\documentclass[12pt,a4paper] {article} 
\usepackage[utf8]{inputenc}
\usepackage{amsmath}
\usepackage{amsfonts}
\usepackage{amssymb}
\usepackage[english]{babel}
\usepackage[left=2cm,right=2cm,top=2cm,bottom=2cm]{geometry}
\usepackage{graphicx}
\usepackage[hidelinks]{hyperref}
\usepackage{lineno}
\usepackage{spverbatim}
\usepackage{mathptmx}
\usepackage[T1]{fontenc}
\usepackage{natbib}

\renewcommand{\vec}[1]{\mathbf{#1}}

\author{Carsten F. Dormann\thanks{Corresponding author; email: \url{carsten.dormann@biom.uni-freiburg.de}} \\ Biometry and Environmental System Analysis \\University of Freiburg, Germany \and Rouven Strauss\\Department of Computer Science \\Technion - Israel Institute of Technology, Haifa, Israel}

\date{\today}

\title{Detecting modules in quantitative bipartite networks: the QuaBiMo algorithm}
\begin{document}
\maketitle
\frenchspacing
\linenumbers

\abstract{
\noindent Ecological networks are often composed of different sub-communities (often referred to as modules). Identifying such modules has the potential to develop a better understanding of the assembly of ecological communities and to investigate functional overlap or specialisation. The most informative form of networks are quantitative or weighted networks. Here we introduce an algorithm to identify modules in quantitative bipartite (or two-mode) networks. It is based on the hierarchical random graphs concept of Clauset \emph{et al.} (2008 Nature 453: 98--101) and is extended to include quantitative information and adapted to work with bipartite graphs. We define the algorithm, which we call QuaBiMo, sketch its performance on simulated data and illustrate its potential usefulness with a case study.
}

\section{Introduction}

The ecological literature is replete with references to interacting groups of species within systems, variously termed compartments \citep{May1973,Pimm1982,Prado2004}, modules \citep{Olesen2007,GarciaDomingo2008,Dupont2009}, cohesive groups \citep{Bascompte2003,DanieliSilva2011,Guimaraes2011} or simply communities \citep{Fortunato2010}. Their attraction, for ecologists, is that they promise a way to simplify the description and understanding of an ecological system, by representing not each and every species, but aggregating their interactions and energy fluxes into a more manageable set of modules \citep[e.g.][]{Allesina2009a}. In the following, we will refer to such aggregated sets of interacting species as `modules'. Their characteristic hallmark is that \emph{within}-module interactions are more prevalent than \emph{between}-module interactions \citep{Newman2003,Newman2004,Fortunato2010}.

In the extreme case, modules are completely separated from each another, and are then typically called compartments \citep{Pimm1982}. This strict definition has seen some relaxation \citep{Dicks2002}, but most recent studies converge on the term module for any identifyable substructure in interaction networks \citep{Prado2004, Lewinsohn2006, Olesen2007, Ings2009, Joppa2009, Cagnolo2010}.

The identification of modules, and the membership of species to modules, has received considerable interest in the physical sciences \citep[as reviewed \emph{in extenso} by][]{Fortunato2010}. Particularly the work of Newman and co-workers \citep[e.g.][]{Newman2003,Newman2004} has practically defined the current paradigm of module definition and identification. Algorithms to identify modules are ``greedy'', i.e.~highly computationally intensive, relying on some way of rearranging module memberships and then quantifying ``modularity'' until a maximal degree of sorting has been achieved \citep{Newman2004a, Clauset2004,Newman2006,Schuetz2008}. The focus of virtually all these algorithms was on unweighted and one-mode networks \citep[see, e.g.,][for a recent examples]{Clauset2008,Lancichinetti2011,Jacobi2012}. Unweighted (or binary or qualitative) refers to the fact that only the presence of a link between species is known, but not its strength \citep{Levins1975,Pimm1982}. One-mode refers to the structure of the community, in which all species are potentially interacting with each other. The typical ecological example is a $n \times n$ food web matrix, in which entries of 1 depict an existing interaction.

\begin{figure}
	\includegraphics[width=\textwidth]{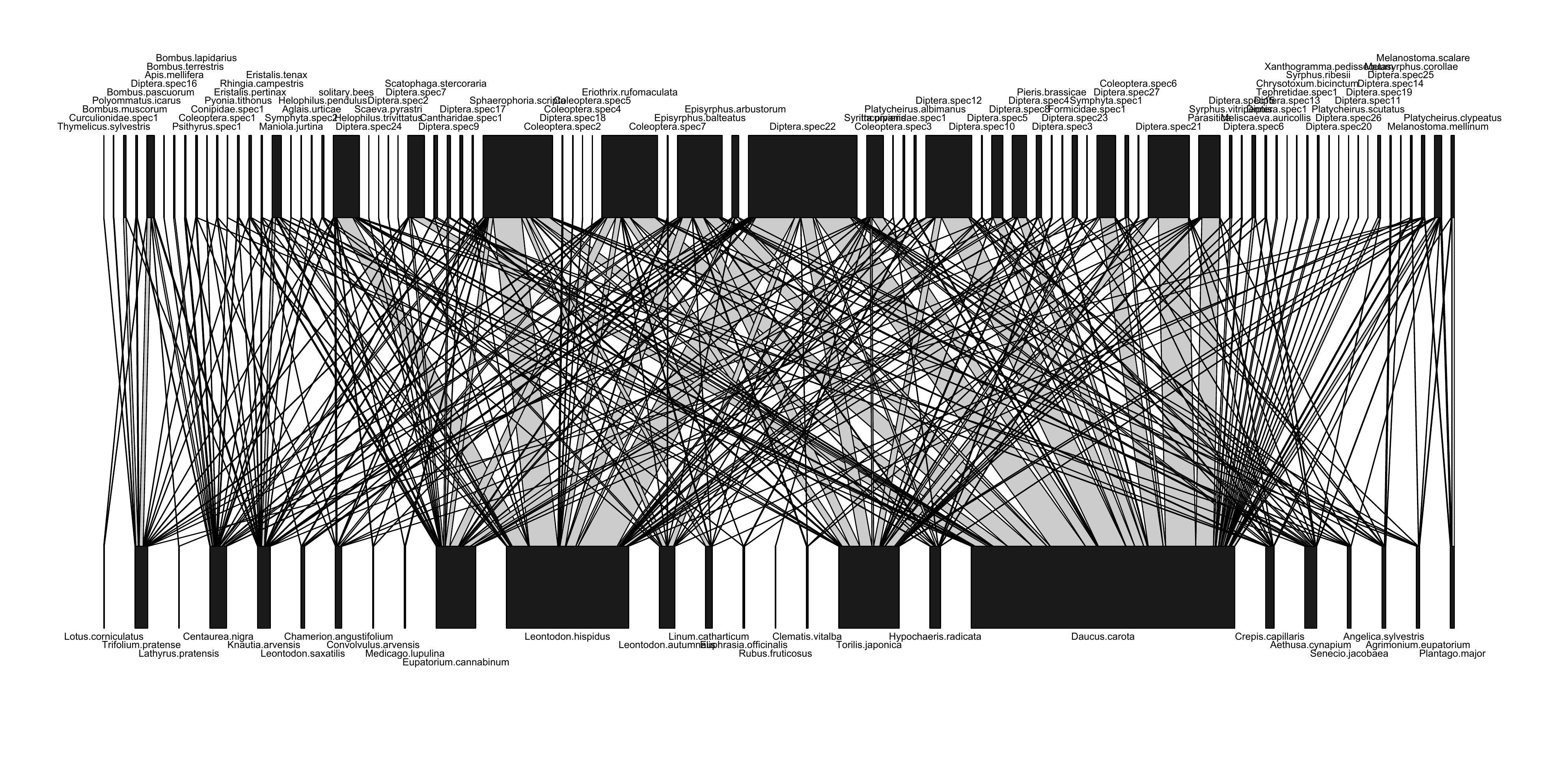}
	\caption{Bipartite graph of a quantitative pollinator-visitation network \citep{Memmott1999}.}
\label{fig:bipartitegraph}
\end{figure}

In recent years, weighted and bipartite interaction networks have become more intensively studies. In a weighted network the link between two species is actually quantified (e.g. by the number of interactions observed or the strength of the interaction inferred from the data). In a bipartite network the species fall into two different groups, which interact with members of the other group, but not within their group. A typical example are pollinator-visitation networks \citep{Vazquez2009}, where pollinators interact with flowers, but flowers do not interact among themselves (see Fig.~\ref{fig:bipartitegraph}). Another well-studied example are host-parasitoid networks \citep[e.g.][]{Morris2004, Tylianakis2007} or seed dispersal networks \citep{Schleuning2012}. 

While popular among ecologists \citep{Bluthgen2010,Schleuning2012,Poisot2012a,Pocock2012}, weighted bipartite graphs are not amenable to any of the existing module-detection algorithms for one-mode networks or for unweighted bipartite networks. Existing software uses only one-mode networks 
or, more precisely, one-mode projections of bipartite networks \citep{Guimera2007, MartinGonzalez2012, Thebault2013}, while other approaches focus on the identification of crucial species through quantifications of their position in the network \citep{Borgatti2006}. This lack of an algorithm to identify modules in quantitative, bipartite networks is particularly problematic, as such networks find their way into conservation ecological considerations \citep{Tylianakis2009} and are the focus of a vibrant field of macroecological research \citep{Ings2009}. Furthermore, from a statistical point of view, weighted networks offer much more information and are less likely to lead to erroneous conclusions about the system \citep{Barber2007, Scotti2007, Bluthgen2010}.

Here we present an algorithm to identify modules (and modules within modules) in weighted bipartite networks. We build on an algorithm provided by \citet{Clauset2008} for unweighted, one-mode networks and the module criterion developed from Newman's one-mode version by \citet{Barber2007}. 

\section{Modularity algorithms}
\begin{figure}
	\hfill
	\includegraphics[width=0.32\textwidth, trim=1cm 4cm 0.5cm 1cm, clip=TRUE]{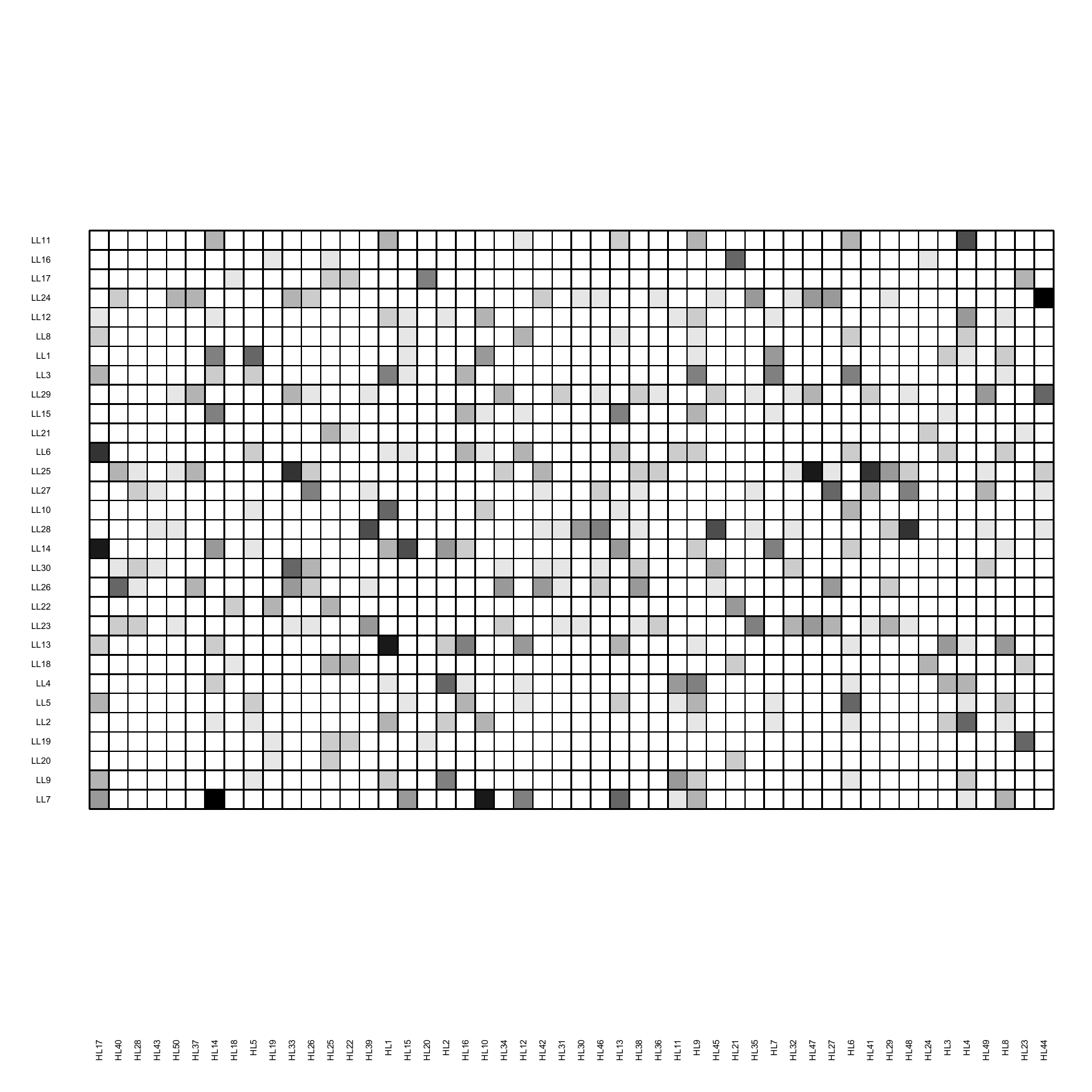}
	\hfill
	\includegraphics[width=0.32\textwidth, trim=1cm 4cm 0.5cm 1cm, clip=TRUE]{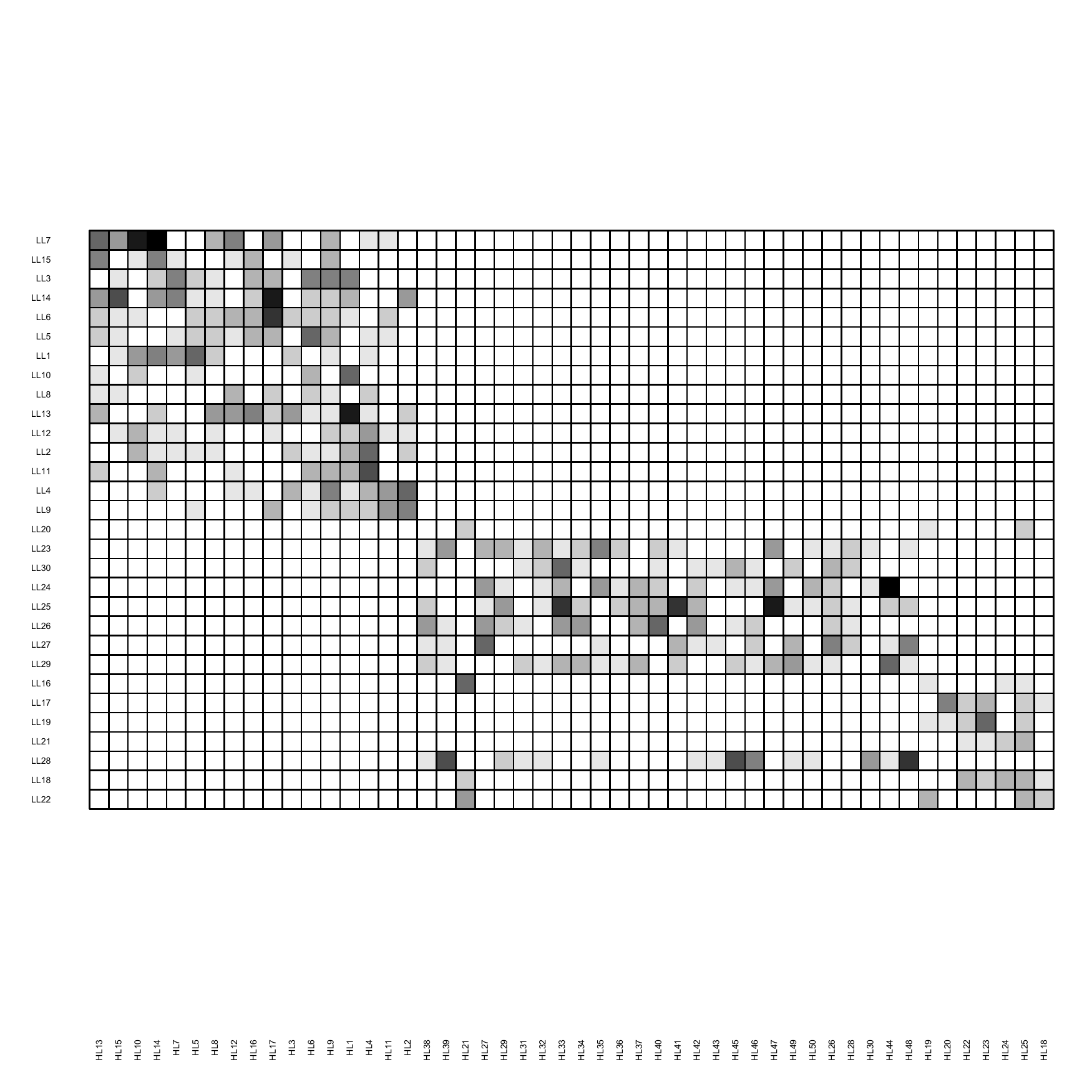}
	\hfill
	\includegraphics[width=0.32\textwidth, trim=1cm 4cm 0.5cm 1cm, clip=TRUE]{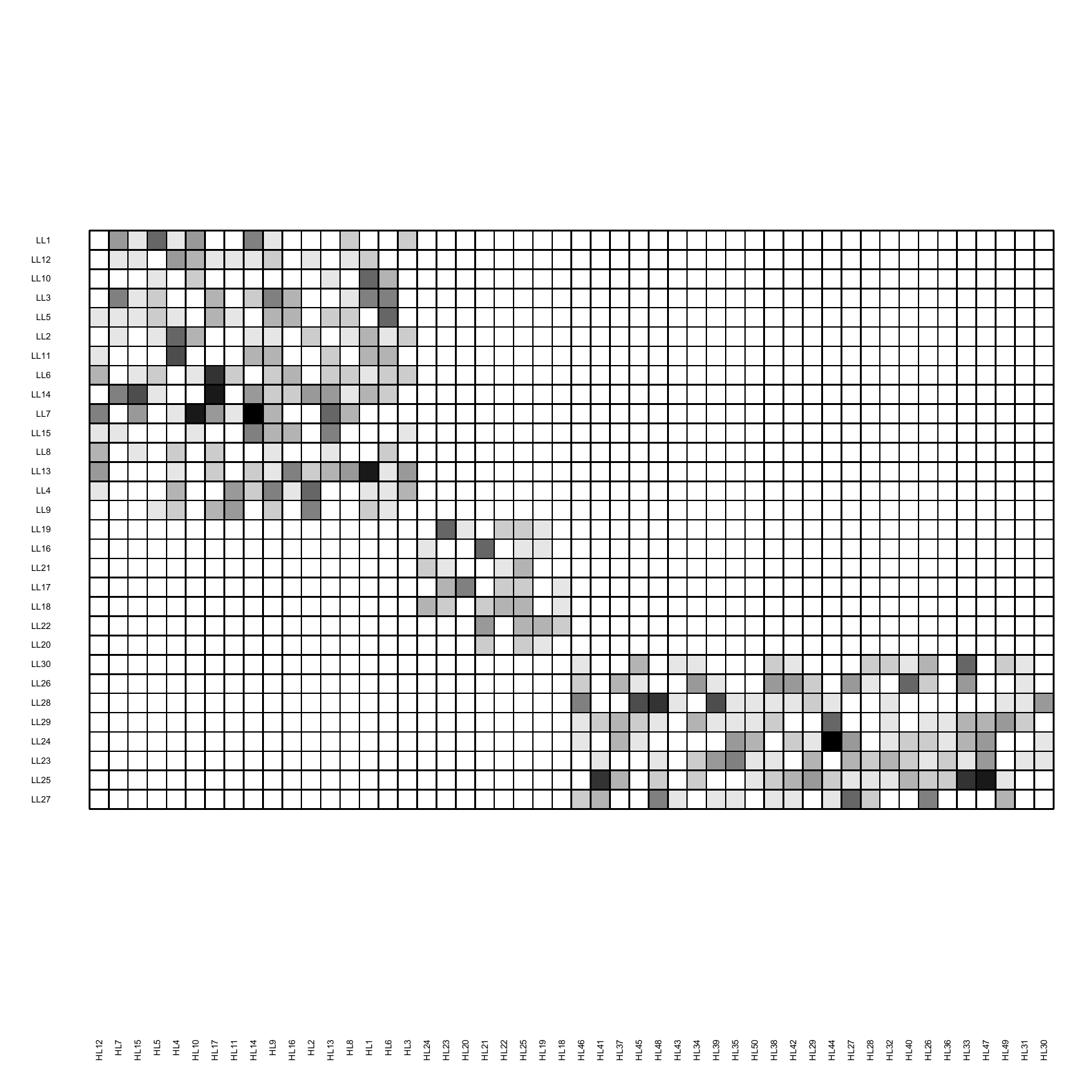}
	\hfill
	\caption{A simulated 3-compartment network in random sequence (left), as sorted by a correspondence analysis (centre) and by the modularity algorithm with default settings (right).}
\label{fig:randomCAsorted}
\end{figure}

Modules can be interpreted as link-rich clusters of species in a community. An alternative to finding and delimiting such modules is to group species by ordination \citep{Lewinsohn2006}. Correspondence analysis (CA) of the adjacency matrix is a simple and fast way to organise species. 
Typically, however, correspondence analysis will not be able to identify modules sufficiently well, even if modules are actually compartments (i.e. perfectly separated: Fig.~\ref{fig:randomCAsorted} left, centre). The QuaBiMo algorithm we present here, can do so, at least in principle (Fig.~\ref{fig:randomCAsorted} right).
If modules are perfectly separated, with no species interacting with species in another module, they are called compartments and will be visible as clearly separated groups of species. It is relatively straightforward to implement a recursive compartment detection function, but compartments are much coarser than modules and not the topic of this publication.

One algorithm proposed and available for detecting modules in bipartite networks is due to \citet[called ``bipart\_w'']{Guimera2007}, which is based on an a one-mode algorithm \citep{Guimera2005}. Their approach differs substantially from a truely bipartite algorithm in that they project the bipartite network into two one-mode networks (one for the higher, one for the lower level) and then proceed identifying the modules for the two levels separately, although they discuss the approach later developed by \citet{Barber2007}. The Guimerà \emph{et al.} approach was used in several ecological applications of modularity \citep{Olesen2007, Dupont2009, Fortuna2010, Carstensen2011, Trojelsgaard2013}, although the algorithm does not allow for the identification of combined modules \citep[as stated in][]{Barber2007, Fortuna2010}. Most recently, \citet{Thebault2013} investigated, through simulations, the ability of three modularity measures \citep[those of ][]{Newman2004, Guimera2007, Barber2007} to identify modules in binary bipartite networks and comes out in support of that of \citet{Guimera2007}.

Finally, \citet{Allesina2009} have proposed an approach for one-mode networks. It identifies ``groups'', rather than modules, which reveal more about the structure of a food web than modules do, since also their relation towards each other emerges from the analysis. Their approach is based on a binary one-mode matrix, however, even when applied to bipartite networks \citep[as was done by][]{MartinGonzalez2012}.

\subsection{QuaBiMo: a Quantitative Bipartite Modularity algorithm}
\subsubsection{Outline}
The new algorithm builds on the Hierarchical Random Graph approach of \citet{Clauset2008}, which builds a graph (i.e.~a dendrogram) of interacting species so that nearby species are more likely to interact. It then randomly swaps branches at any level and evaluates whether the new graph is more likely than the previous one, recording and updating the best graph. The swapping is a Simulated Annealing-Monte Carlo approach, i.e. sometimes a worse graph is chosen as the starting point for the next swap, thereby avoiding being trapped in a local maximum. Each node of the graph contains the information of whether it is part of a module, so that the graph can be transgressed top-down to identify modules.

Our modifications consists of (a) allowing branches between species to be weighted by the number of interactions observed between them, thereby making the algorithm quantitative; and (b) taking into account that species in one group can only interact with species in the other group, rather than the one-mode network the algorithm was initially developed for. Taken together, our algorithm computes modules in weighted, bipartite networks, based on a hierarchical representation of species link weights and optimal allocation to modules.

\subsubsection{Terminology}
A graph $G=(V,E)$ denotes a set of vertices $v \in V$ connected by edges $e \in E$. An edge $e$ connects two nodes, thus $e=c(v_i, v_j)$, where $v_i \in V \wedge v_j \in V$. $G$ is a weighted (= quantitative) graph if each edge $e$ has a weight $w \in W$ associated with it ($w\subseteq \mathbb{R}^+$). We normalise edge weights so that $\sum_{w \in W}w=1$. (For binary graphs $w=1/|E|$ for all existing edges, where $|.|$ symbolises the number of elements.)

For bipartite graphs, the vertices $V$ are of two non-overlapping subsets, $V_H$ and $V_L$ (higher and lower level), such that $V_H \cap V_L = \emptyset$ and for all edges the connected vertices are in different subsets: $v_i \in V_H \iff v_j \in V_L$ ($\iff$ symbolises equivalence, i.e.~if we know $v_i$ is in $V_H$, $v_j$  \emph{must} be in $V_L$, and vice versa).

A graph can be represented as a dendrogram $D$, i.e.~a binary tree with the vertices of the graph $G$ being the tips (or leaves) of the dendrogram $D$. Thus, any internal split (or vertex) of $D$ defines a subset of $G$. The idea of the algorithm is now to find internal vertices of $D$ so that the subset it defines is a module.

\subsubsection{Goal function}

The algorithm has to divide $G$ into a set of modules $M$ such that 
\begin{enumerate}
	\item each module $m \in M$ is a connected subgraph of $G$. (This means each species has to have a partner.)
	\item each vertex $v$ belongs to exactly one module $m$. (The uniqueness requirement.)
	\item edge weights within a module are higher than edge weights outside modules. (The modularity definition.)
\end{enumerate}
To specify point 3 above, \citet{Barber2007} has defined modularity for weighted bipartite networks as 
\begin{equation}\label{eq:Q}
Q = \frac{1}{2N} \sum_{ij}{\left(A_{ij}-K_{ij}\right)}\delta(m_i, m_j)
\end{equation} 
where $N$ is the total number of observed interactions in the network and $A_{ij}$ is the normalised observed number of interactions between $i$ and $j$, i.e. the edge matrix $\vec{E}$. The expected value, based on an appropriate null model, is given in the matrix $\vec{K}$ (see below). (Without normalisation, $\vec{A}$ and $\vec{K}$ represent the adjacency matrix and the null model matrix, respectively.) 
The module to which a species $i$ or $j$ is assigned is $m_i$, $m_j$. The indicator function $\delta(m_i, m_j)= 1$ if $m_i = m_j$ and $0$ if $m_i \neq m_j$. $Q$ ranges from $0$, which means the community has no more links within modules than expected by chance, to a maximum value of $1$. The higher $Q$, the more do the data support the division of a network into modules.

One crucial point of our modifications of the original algorithm was to assign an indicator value to each dendrogram vertex to label it as being within a module, or not. To do so, we have to compute the expected value for each value of $A_{ij}$ in order to be able to evaluate whether the observed value is lower or higher (the term over which eqn.~\ref{eq:Q} sums). This step is not required if edges are unweighted, since then the expectation will always be the same. For weighted edges however, we would expect the edge $e_{ij}$ connecting two nodes $i$ and $j$ representing abundant species to have a high value of $w_{ij}$. Similarly, nodes representing rare species could be expected to have low edge weights.

Thus, at every vertex of the tree, the algorithm assembles the module defined by the vertex' position (i.e. including all leaves on its branches) and computes the expectation matrix $\vec{K}$ based on the cross product of marginal totals of all species in the module $\vec{A}_{.j}$ and $\vec{A}_{i.}$, divided by the sum of the number of observed interactions in that module: $\vec{K}=\vec{A}_{.j}^T \vec{A}_{i.}$. Since we normalised all edge weight to sum to $1$, $\vec{K}$ is actually a probability matrix. If the vertex gives rise to a module, i.e. if $\sum_{ij \in m}{\left(A_{ij}-K_{ij}\right)} > 0$, this vertex is labelled as a module. We can now sum the contributions of all vertices and modules according to eqn.~\ref{eq:Q} to compute to total modularity of graph $G$. For a formal description of this part of the algorithm, please see appendix A.

\subsubsection{Swapping}
The algorithm starts with a random dendrogram, where modularity $Q$ is likely to be very low. Through random swapping of branches and their optimisation, $Q$ increases during a Simulated Annealing procedure. The algorithm stops when a pre-defined number of swaps did not further increase the value of $Q$.

\begin{figure}
\centering
\includegraphics[width=.7\textwidth]{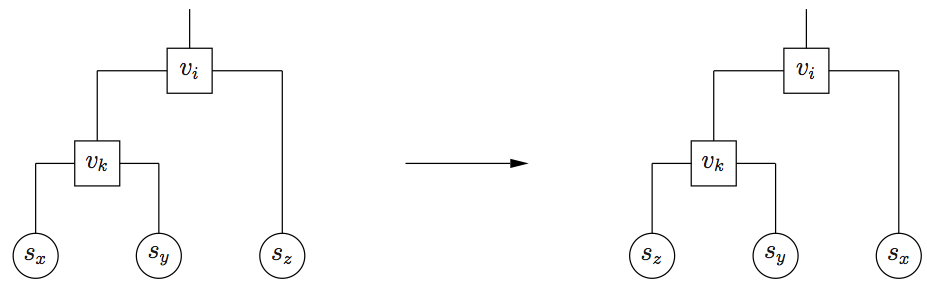}
\medskip
\includegraphics[width=.7\textwidth]{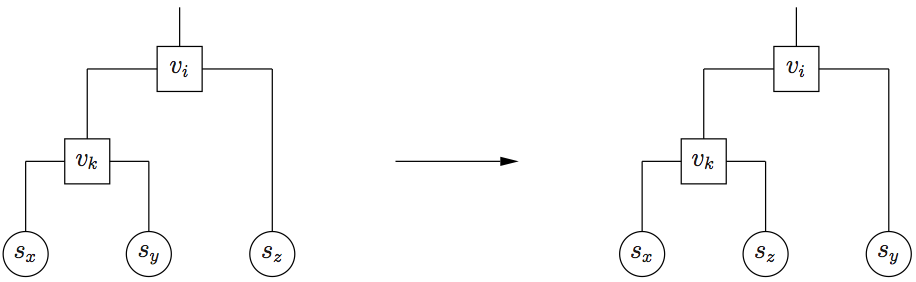}
\caption{Two possible moves in the swapping of randomly selected vertices $v_i$ and $v_k$. The leaves depicted here could be actual leaves or, more often, sub-tree and are hence labelled $s$. The algorithm randomly chooses one of these two possible new configurations.}
\label{fig:swaps}
\end{figure}

Random swaps are implemented as exchange of two randomly selected vertices in the dendrogram, subject to the following constraint (Fig.~\ref{fig:swaps}). The vertex to be swapped cannot be a leaf. Since terminal vertices always connect leaves from the two bipartitions $V_i$ and $V_j$, thus representing an interaction, they can be swapped, while their leaves cannot.

After each swap, the modularity of the entire dendrogram is re-computed (for computational efficiency only those parts affected by the swap). If the new configuration has a higher value of $Q$ it is stored and becomes the new best dendrogram, otherwise the previous configuration will be used as the starting point for the next swap. A worse configuration is accepted with the probability $p < e^{\frac{\delta Q}{T}}$, where $\delta Q$ is the change in modularity from the last configuration to the new one and $T$ is the current temperature of the Simulated Annealing algorithm. We observed that the algorithm converges notably faster if the temperature is not decreased monotonously, but rather set back to the average temperature at which an increase in $Q$ occurs. This is also a better approach in our case, since we do not know, a priori, how many steps the algorithm will take, or which value of $Q$ can be obtained.

Since the hierarchical dendrogram is computed through iterative proposing, evaluation and rejecting dendrogram structure in a Markov Chain Monte Carlo approach, \citet{Clauset2008}'s, and hence our, algorithm cannot guarantee finding the optimal module configuration. Since the algorithm is coded in C++, even billions of MCMC swaps are feasible in a few minutes, yielding reasonable results for typically sized ecological networks (see below) at acceptable handling time. For large networks, this algorithm can run for hours to days. See section \ref{sec:session} for an example session on how to employ the algorithm through R \citep{RTeam}.

\subsection{Output \& nested modules}
The algorithm returns an object identifying modules and sequence-vectors for species, as well as a re-order network ready for visualisation of modules and the modularity $Q$.

QuaBiMo can be invoked recursively, searching for modules within modules. While such nested modules become ever smaller and are thus ever faster to detect, there are plenty of them and hence nesting will typically dramatically prolong the search for patterns.


\section{Evaluation of the algorithm}

The detection of modules has theoretical limits related to the number of between-module links present, the sparceness of the network matrix and the size of the network \citep[e.g.][]{Fortunato2007, Lancichinetti2011, Lancichinetti2010}. In the following paragraphs we evaluate the QuaBiMo-algorithm for different simulated networks typical in size and noise for pollination networks. There is no technical reason why the algorithm should not work for much larger networks, too, given enough time for computing a large number of dendrogram configurations. Such an evaluation is outside the scope of this study.

\subsection{Simulations to investigate algorithm sensitivity and specificity for noisy network data} 
\begin{figure}
\includegraphics[width=\textwidth]{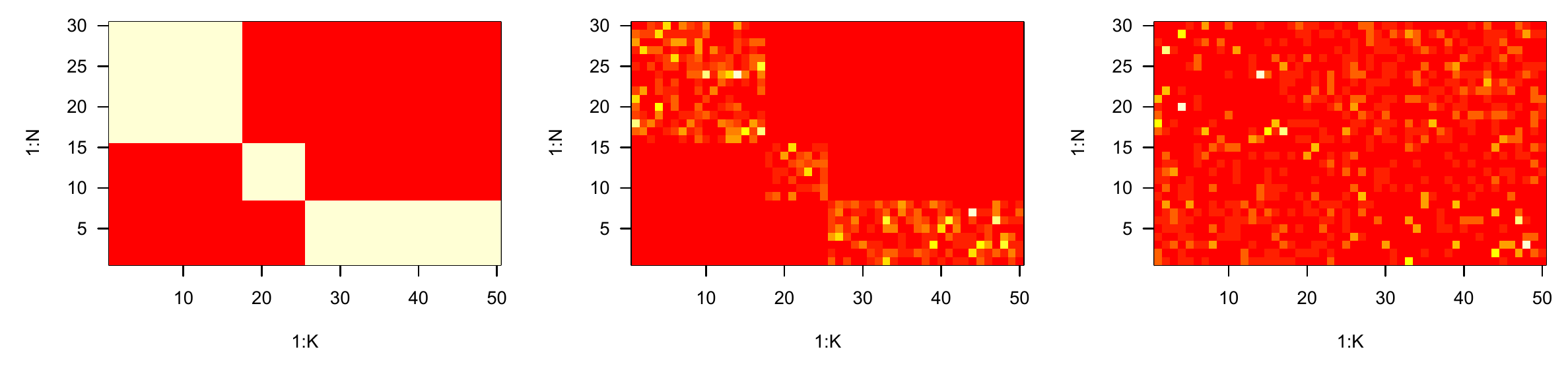}
\caption{Network simulation starts by defining the modules (left), then allocating to all links a number of interactions drawn from a negative binomial distribution (centre) and finally removing interactions in a module and placing them outside (right). High levels of noise, as shown here, yield poorly defined modules. Cells with a value of $0$ are shown in red.}
\label{fig:noisesimu}
\end{figure}

We analysed simulated networks of different noisiness to evaluate the performance of the modularity algorithm. We would expect that modules become unidentifiable when the proportion of links within modules becomes as low as between modules. We hence simulate networks with increasing degree of noise by moving, randomly, interactions from within a module to a random position in the adjacency matrix not included in any module (Fig.~\ref{fig:noisesimu}).
We simulated two sizes of networks (30 $\times$ 50 and 100 $\times$ 400), two levels of filling (achieved through setting the parameter ``size'' of the negative binomial distribution to 0.1,``low'', or 1, ``high'' ), and two levels of modularisation (3 and 10 modules). Each combination was evaluated for seven noise levels (0, 0.05, 0.1, 0.2, 0.3, 0.4, 0.5) and replicated 15 times, yielding 840 different networks. Replicates differ in the size, position of modules and number of interactions per link. Sizes were maintained at the same two levels.

Networks were simulated in three steps. First, we defined the size of the matrix and position and size of the modules. This initial network is a matrix of $0$s except for all interactions in a module, which is thus identified by a block of $1$s. Then, second, we drew actual interactions for each link of a module from a strongly skewed negative binomial distribution (with size $=0.05$ and $\mu=2$), removed 80\% (high filling) or $40\%$ (low filling) of $0$-values,  and then replaced the initial $1$s of the module blocks by these random values. Accordingly, the modules had a connectance (= filling) of less than $100\%$. Higher filling of modules generally increases performance. Third, we randomly drew a proportion of interactions from the module and moved it to randomly selected columns and rows of these species outside the module. Thereby we effectively added noise to the network data. There is an upper limit to the third step, where modules become ill-defined. That is the case when the number of interactions outside modules is as high as inside.

We ran the QuaBiMo algorithm five times on each network, saving the result with the highest modularity. This was more efficient in finding a good module configuration than running the algorithm for much longer. For comparison, we also ran the algorithm on a binarised version of the data. The code for simulations and analysis is available in appendix B; runtime for the simulatios was approximately two weeks on a standard desktop computer with 32 GB RAM.

Congruence between the original assignment to modules and the one identified by the algorithm was assessed by means of a confusion matrix. Each link existing in the simulated data was classified as correctly belonging to a module, falsely assigned to a module, falsely not assigned to a module, or correctly not assigned to a module. The confusion matrix was then summarised as sensitivity, specificity and accuracy.

\subsection{Simulation results: modularity \emph{Q} in binary and weighted networks}
\begin{figure}
	\centering
	\hfill
	\includegraphics[width=0.45\textwidth]{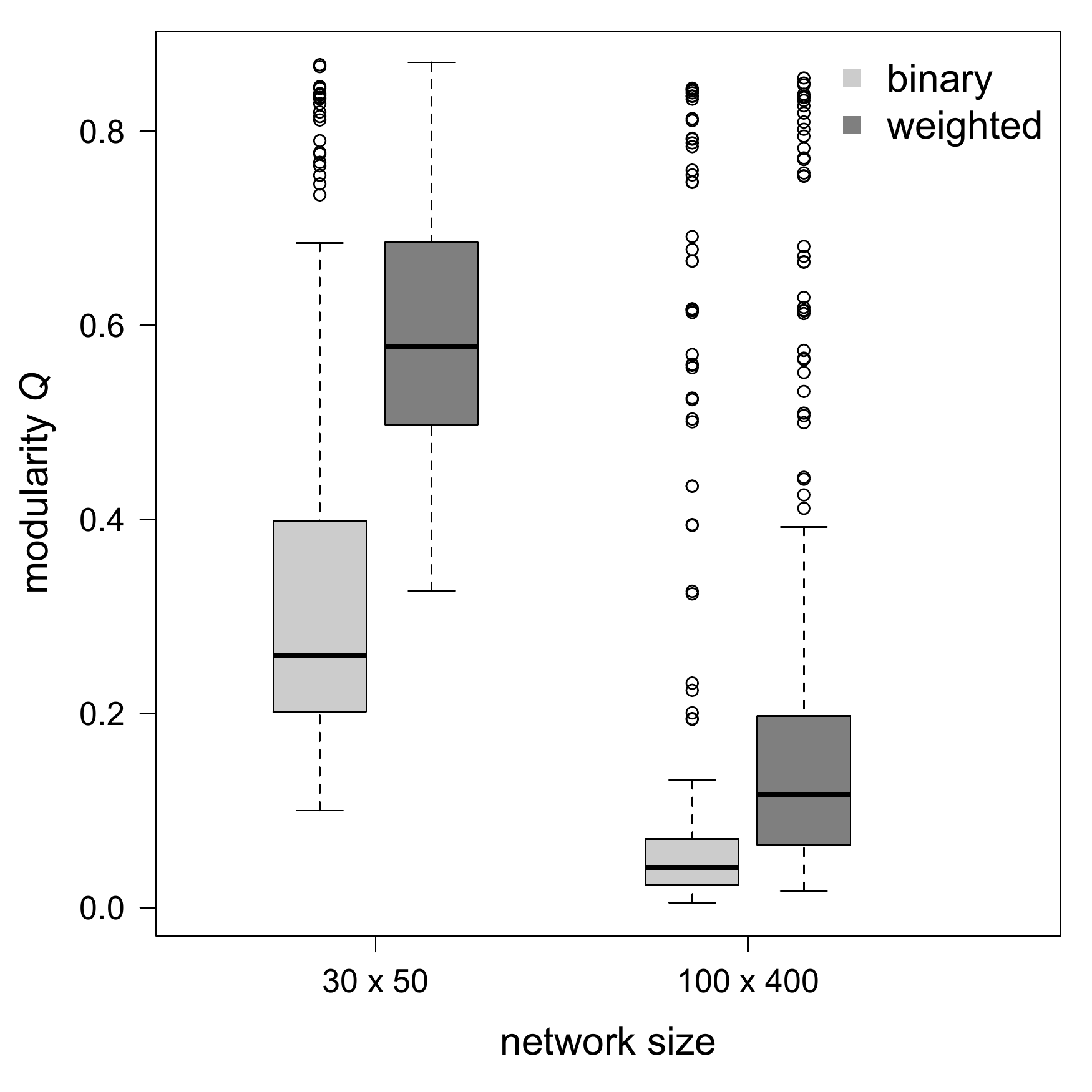}
	\hfill
	\includegraphics[width=0.45\textwidth]{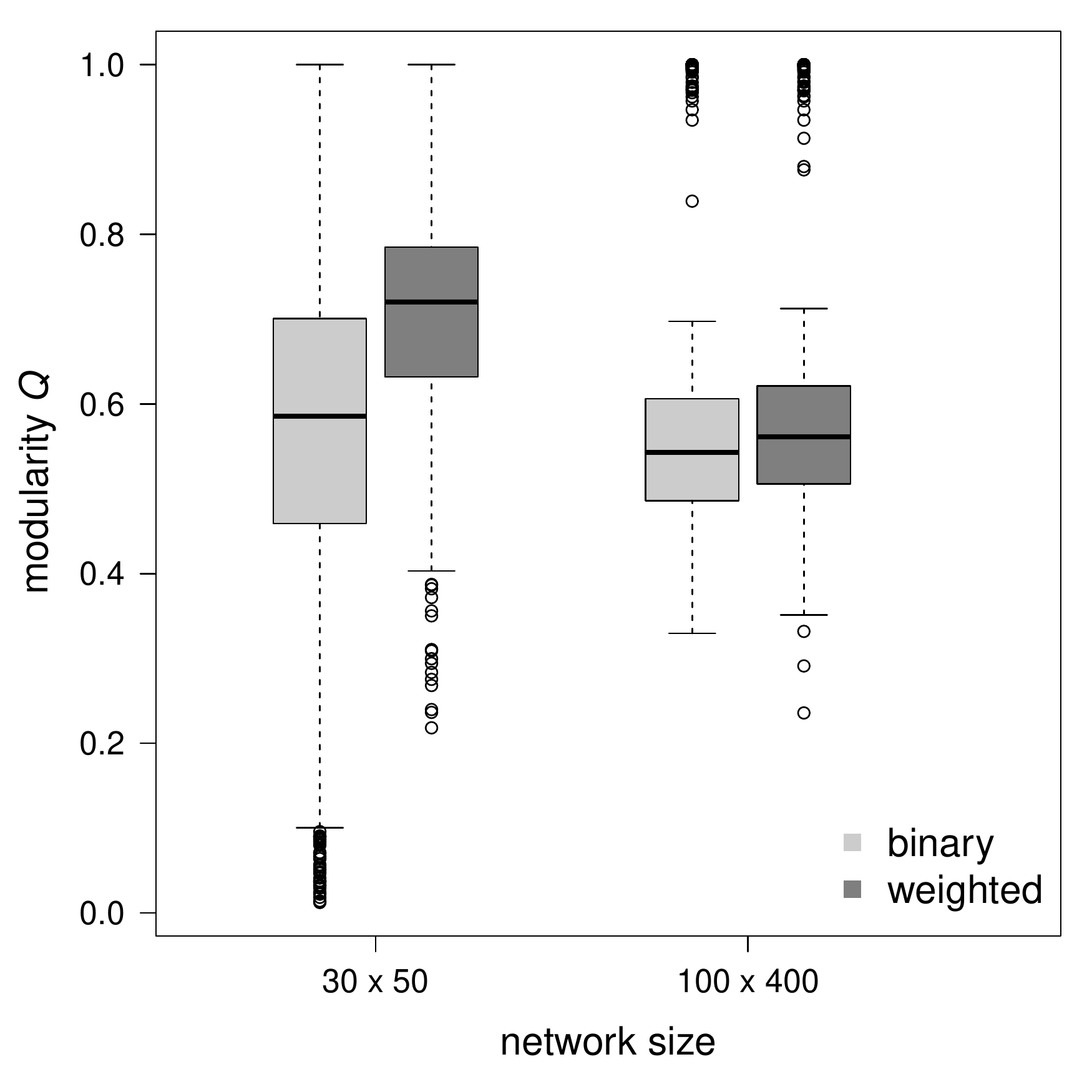}
	\hfill
	\caption{Quality of modularity detection (left: $Q$; right: overall accuracy) depends on network size and type of information (binary or weighted).}
\label{fig:QsizeXbinary}
\end{figure}

Modularity $Q$ was strongly dependent on network size, the amount of noise added and the number of modules (Table~\ref{tab:Qanova}). Most importantly, however, our quantitative approach strongly improved on modularity based on binary data, particularly for large networks (Fig.~\ref{fig:QsizeXbinary}). Deterioration of the module detection with increasing network size could possibly be compensated for by increasing the number of swaps before terminating the search (see example session below). The loss of skill with increasing noise (Fig.~\ref{fig:QsizeXbinary}, right) cannot be alleviated. Here the ability of QuaBiMo to use not only the binary but the weighted link information is already a dramatic improvement.
\begin{table}
\caption{Effect of different simulation parameters on modularity $Q$ and overall accuracy. Sum of squares and $F$-value can be taken as a measure of how strongly these parameters effect modularity. No significances are given since a test of an effect is nonsensical for simulations. Information refers to binary vs. weighted networks. `Noise' has seven levels but was analysed as continuous variable.}
\label{tab:Qanova}
\centering
\begin{tabular}{lrrr}\hline
\textbf{Modularity $Q$}& df & sum of squares & $F$ value \\\hline
noise         					 	&   1    & 17.78  & 1024 \\
size             				  	  &   1 	& 38.24 & 2204  \\
fill               						&  1  & 1.06   & 61 \\
no.of.modules 			&  1 &  6.02    & 346  \\
information					& 1  & 11.65 & 671 \\
noise:size						& 1  & 0.30   & 18 \\
noise:no.of.modules & 1  &  2.01 & 116 \\
size:fill				             & 1  & 0.57 &  33\\
size:no.of.modules	& 1  & 0.42   & 24\\
size:information  	  & 1 & 4,27 & 246 \\
Residuals          		 	   & 1635 & 28.37 & \\\hline
\textbf{Overall accuracy}& df & sum of squares & $F$ value \\\hline
noise         					 	&   1    &   7.59 & 296 \\
size             				  	  &   1 	& 0.24 & 9  \\
fill               						&  1  &  0.74 & 29 \\
no.of.modules 			&  1 &  2.30  & 90  \\
information			  	   & 1  & 2.60  & 102 \\
noise:size					  & 1 & 0.35 & 14 \\
noise:fill						& 1 & 0.62 & 24\\
noise:no.of.modules & 1 & 0.46 & 18 \\
noise:information & 1 & 2.33 & 91 \\
size:fill							& 1  &  0.24 & 9 \\
size:no.of.modules  & 1  & 1.32  & 52 \\
size:information 	  & 1  & 1.81 &  71 \\
no.of.modules:information  	& 1  & 0.87  & 34\\
Residuals          		 	   & 1632 & 41.86 & \\\hline
\end{tabular}
\end{table}

In the following paragraphs, we shall only be looking at the results for the weighted networks, since that is the explicit focus of the QuaBiMo algorithm.

\subsection{Simulation results: modularity \emph{Q} }
Modularity $Q$ and overall accuracy were affected very similarly by network size, noise and the number of modules (Table~\ref{tab:Q}).
\begin{table}
\caption{Effect of different simulation parameters on modularity $Q$ for weighted networks. Sum of squares and $F$-value can be taken as a measure of how strongly these parameters effect modularity.}
\label{tab:Q}
\centering
\begin{tabular}{lrrr}\hline
									& df & sum of squares & $F$ value \\\hline
noise         					 	&   1    & 7.56  & 584 \\
size             				  	  &   1 	& 33.64 & 2597  \\
fill               						&  1  &  0.59 & 45 \\
no.of.modules 			&  1 &   3.76   & 290  \\
noise:size						& 1 & 0.40 & 31\\
noise:no.of.modules & 1 & 0.55 & 43 \\
size:fill							& 1  &  0.23  & 18 \\
size:no.of.modules  & 1  &  0.12 & 9 \\
Residuals          		 	   & 809 & 10.48 & \\\hline

\end{tabular}
\end{table}
\begin{figure}
	\centering
	\hfill
	\includegraphics[width=0.45\textwidth]{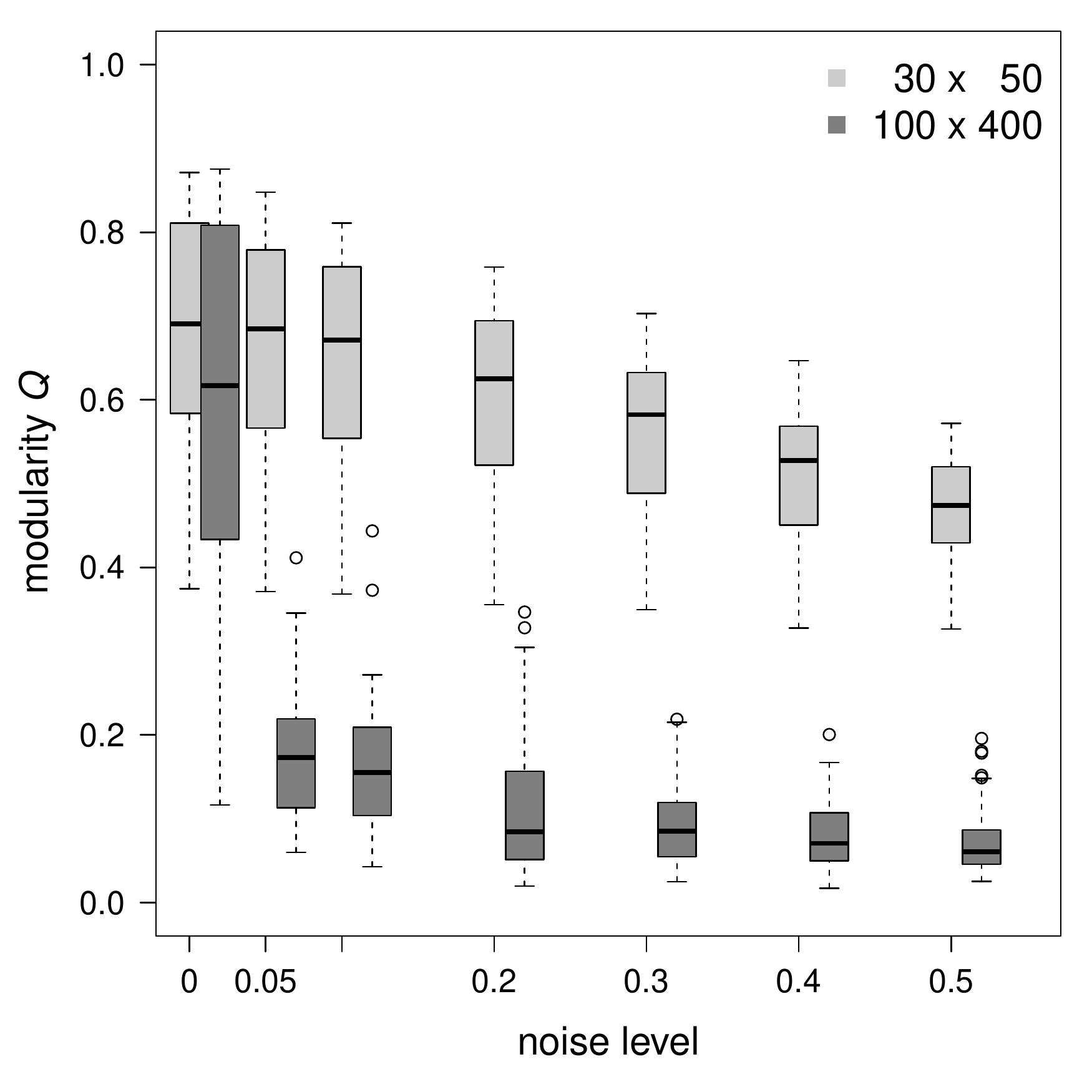}
	\hfill
	\includegraphics[width=0.45\textwidth]{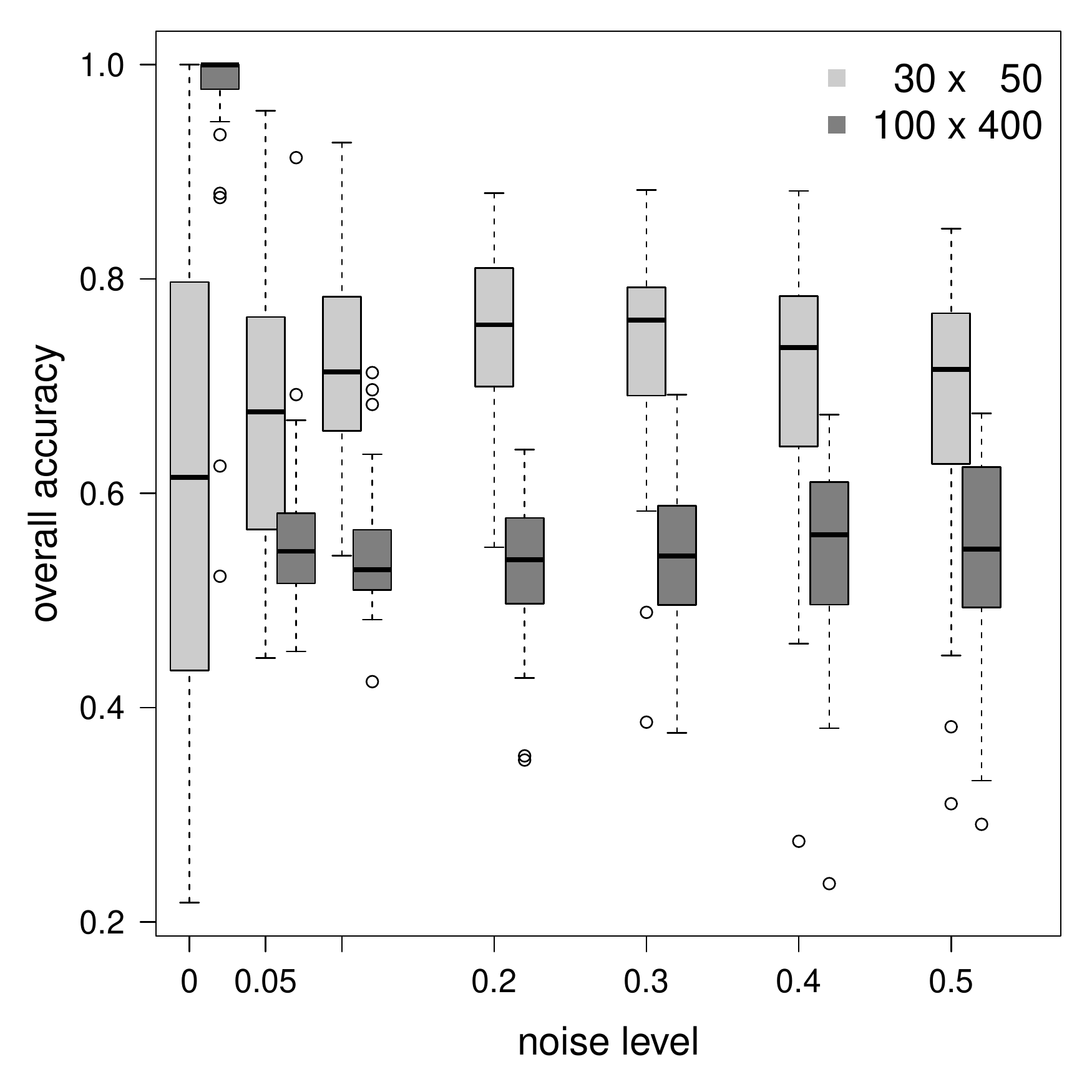}
	\hfill
	\caption{Effect of noise and network size on modularity $Q$ (left) and overall accuracy (left).}
\label{fig:QandAccuracy}
\end{figure}
The most prominent effects were those of size, noise and their interaction, depicted for $Q$ and overall accuracy in Fig.~\ref{fig:QandAccuracy}. Evidently, larger networks are more difficult to modularise, as are those with a higher level of noise.

\subsection{Simulation results: classification accuracy}
While modularity $Q$ gives an indication of how well observed links could be grouped into modules (with a value of 1 indicating that all links are within and none between modules), we can also quantify the algorithm's accuracy based on a confusion table. Overall accuracy (= correct classification rate) is the proportion of all links correctly placed, i.e. (number of links correctly placed into modules + number of links correctly placed between modules)/total number of links. Since the purpose of the algorithm is the use of weighted network data, we here only present results for the weighted and not for the binary networks.

The overall accuracy of module detection decreased with increasing noise levels (Table~\ref{tab:accuracyanova}), an effect more pronounced for large networks than for small ones (Fig.~\ref{fig:QandAccuracy} right). Again, this interaction probably could have been reduced if more steps until termination were allowed for the larger networks.
\begin{table}
\caption{Effect of different simulation parameters on module identification accuracy (weighted networks only). }
\label{tab:accuracyanova}
\centering
\begin{tabular}{lrrr}\hline
										  & df & sum of squares & $F$ value \\\hline
noise         					 	&   1    & 0.74 & 41 \\
size             				  	  &   1 	& 1.65 & 90  \\
fill               						&  1  & 0.84 & 46 \\
no.of.modules 			&  1 &  0.18 & 10  \\
noise:size						& 1 & 1.95 & 106\\
noise:fill						  & 1 & 0.14 & 8 \\
noise:no.of.modules & 1 & 0.11 & 6 \\
size:fill							& 1  &  0.32  & 17 \\
size:no.of.modules  & 1  &  0.17 & 9 \\
Residuals          		 	   & 808 & 14.80 & \\\hline
\end{tabular}
\end{table}

\subsection{Simulation results: sensitivity and specificity}
Classification accuracy has two elements: the correct classification of all module links as belonging to modules (sensitivity) and the correct identification of between-module links as \emph{not} belonging into modules (specificity). For the detection of patterns in networks high sensitivity is desirable, although this may inflate type II errors (i.e. we may identify modules that do not really exist). High specificity indicates that links allocated into modules are indeed correct, but possibly at the expense of not allocating many links to modules overall (leading to inflated type I errors).

\begin{figure}
	\centering
	\hfill
	\includegraphics[width=0.45\textwidth]{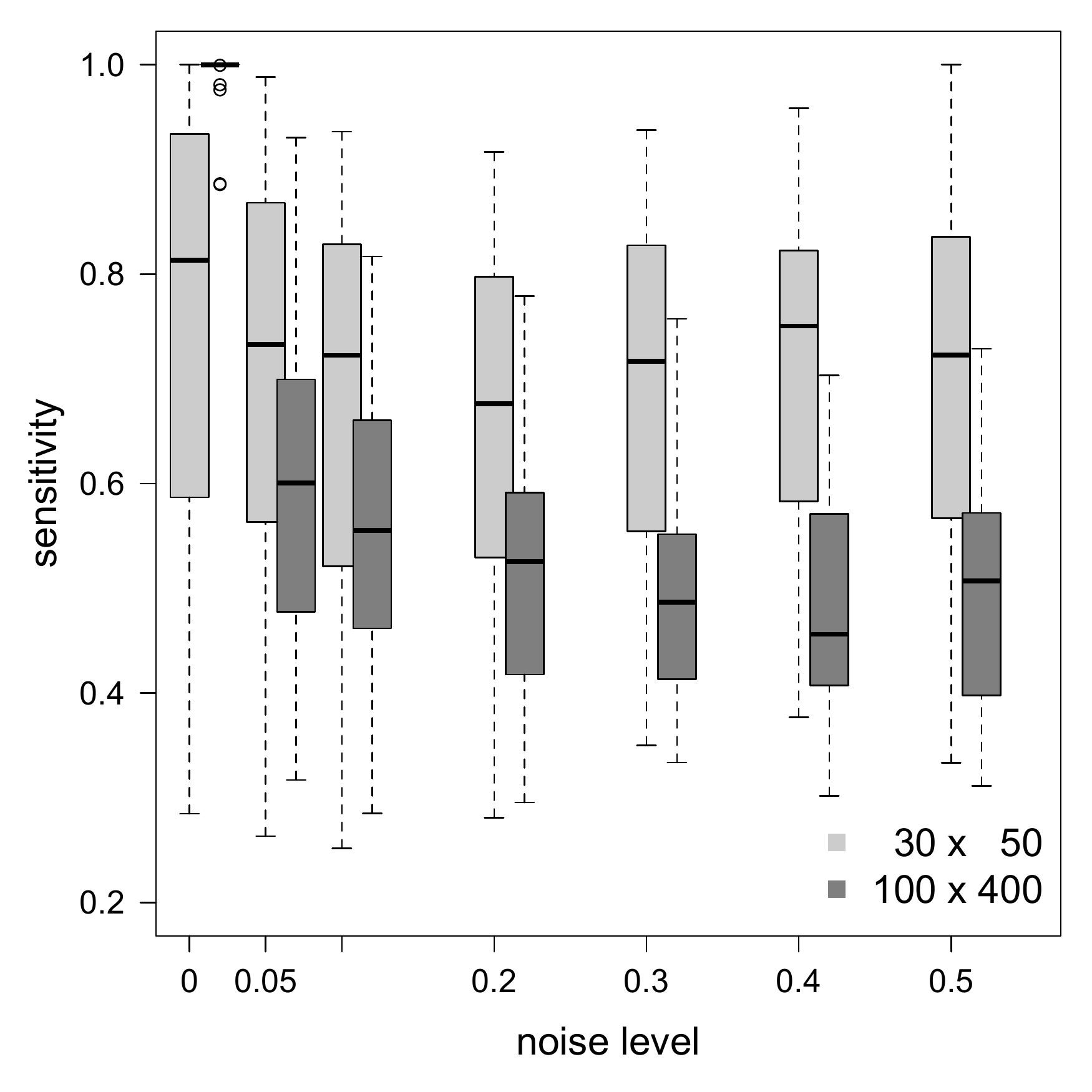}
	\hfill
	\includegraphics[width=0.45\textwidth]{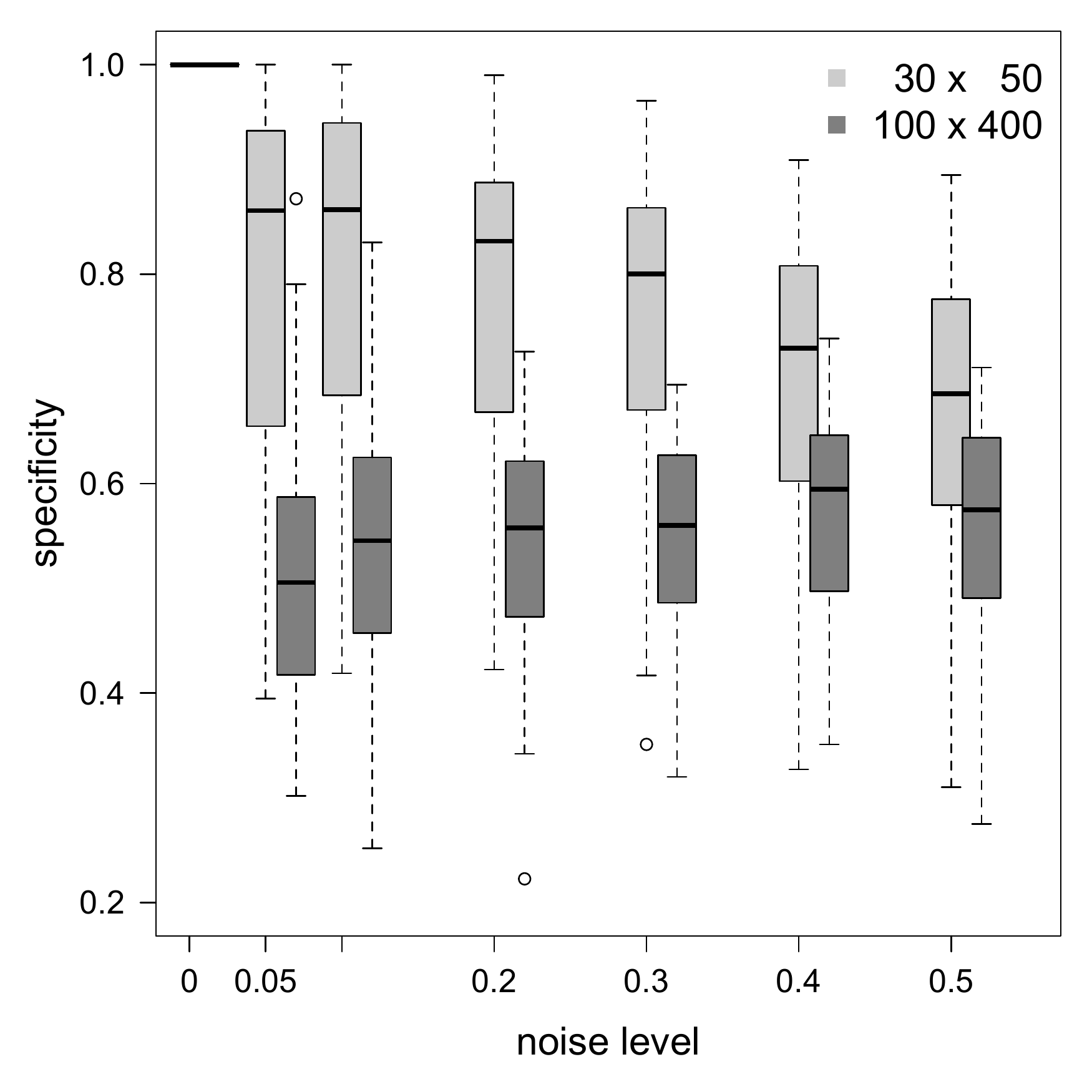}
	\hfill
	\caption{Effect of noise and network size on sensitivity (right) and specificity (left) of the classification of links into modules.}
\label{fig:sensspec}
\end{figure}

Sensitivity and specificity of the QuaBiMo-algorithm were driven by the same factors as overall accuracy (Table \ref{tab:SensSpec}). Increasing noise levels reduced both sensitivity and specificity, as did larger networks (Fig.~\ref{fig:sensspec}).
\begin{table}
\caption{Effect of different simulation parameters on sensitivity and specificity of module identification (weighted networks only).}
\label{tab:SensSpec}
\centering
\begin{tabular}{lrrr}\hline
\textbf{Sensitivity} & df & sum of squares & $F$-value\\\hline
noise     		&    1  &  8.48 & 278 \\
size        	   &   1  &  3.34 &  109  \\
fill 				 & 1    & 0.46 & 15 \\
no.of.modules & 1 & 3.01 & 99 \\ 
noise:size  &   1 & 0.27 & 9\\
noise:fill 	  & 1  & 0.65 & 21 \\
size:fill      & 1  & 0.81 & 27\\
Residuals  & 810 & 24.75 & \\\hline
\textbf{Specificitiy} & df & sum of squares & $F$-value\\\hline
noise     		&  1 &  5.84 &  308 \\
size        	   & 1  & 11.73 & 618  \\
fill 				&  1 & 0.48 & 25 \\
no.of.modules  & 1 & 0.24  & 13\\
Residuals  & 813 & 15.44 & \\\hline
\end{tabular}
\end{table}

\section{Identifying modules - an example session}
The QuaBiMo-algorithm is implemented in C++ and is made available through the open source R-package \texttt{bipartite} \citep{Dormann2009}. The most important function is \texttt{computeModules}, which takes three arguments: the matrix representing the bipartite network data (``web''), a specification of how many MCMC moves should yield no improvement before the algorithm stops (``steps'', with default $1E6$) and a logical switch for computing nested modules (``deep'', defaulting to FALSE). The number of steps should be adapted to the size of the network (see previous sections). We found that $Q$ levels off very soon, once the default of one million is exceeded. However, we have not extensively trialled this setting for networks larger than that used below.

\begin{figure}
	\includegraphics[width=\textwidth]{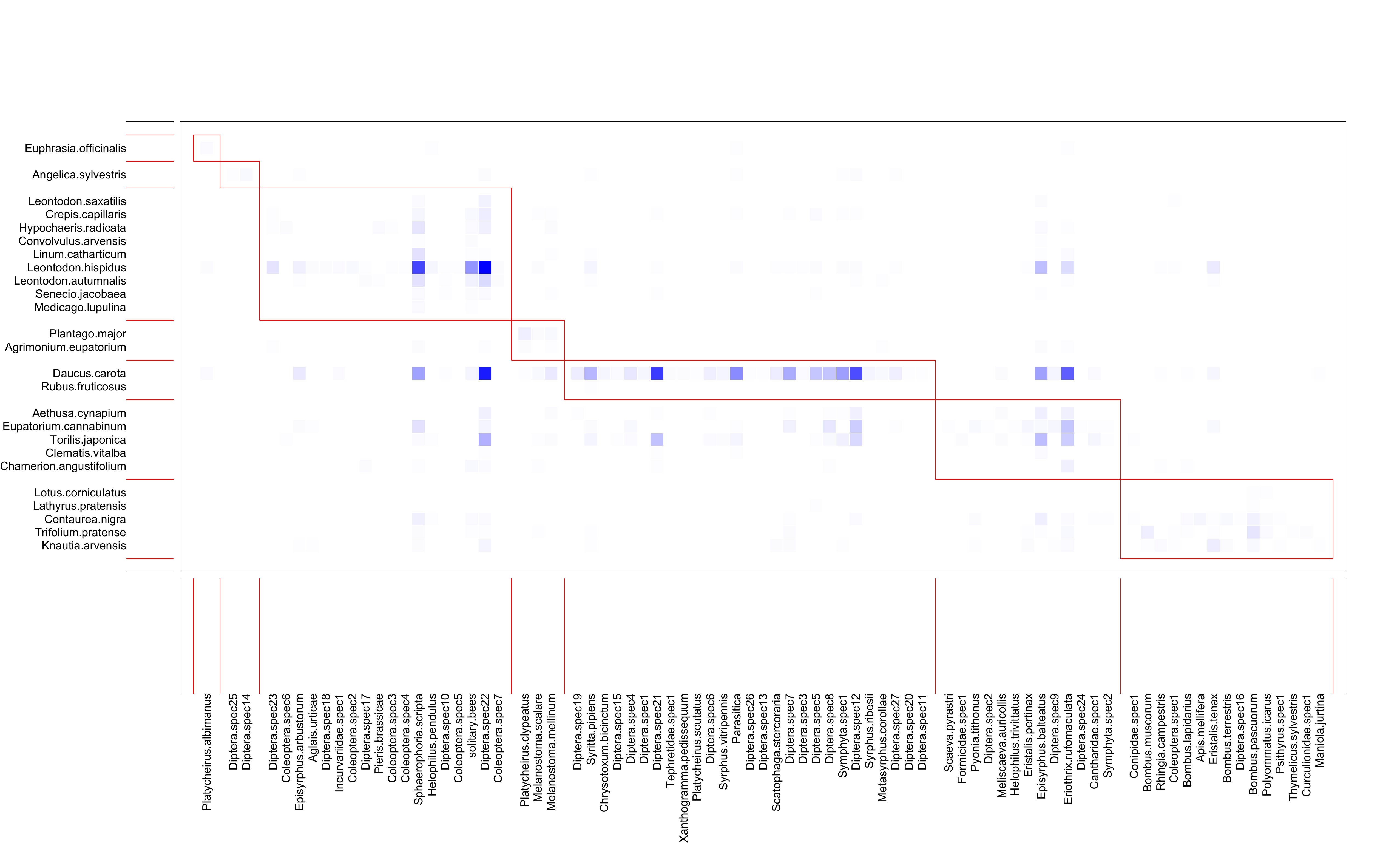}
	\includegraphics[width=\textwidth]{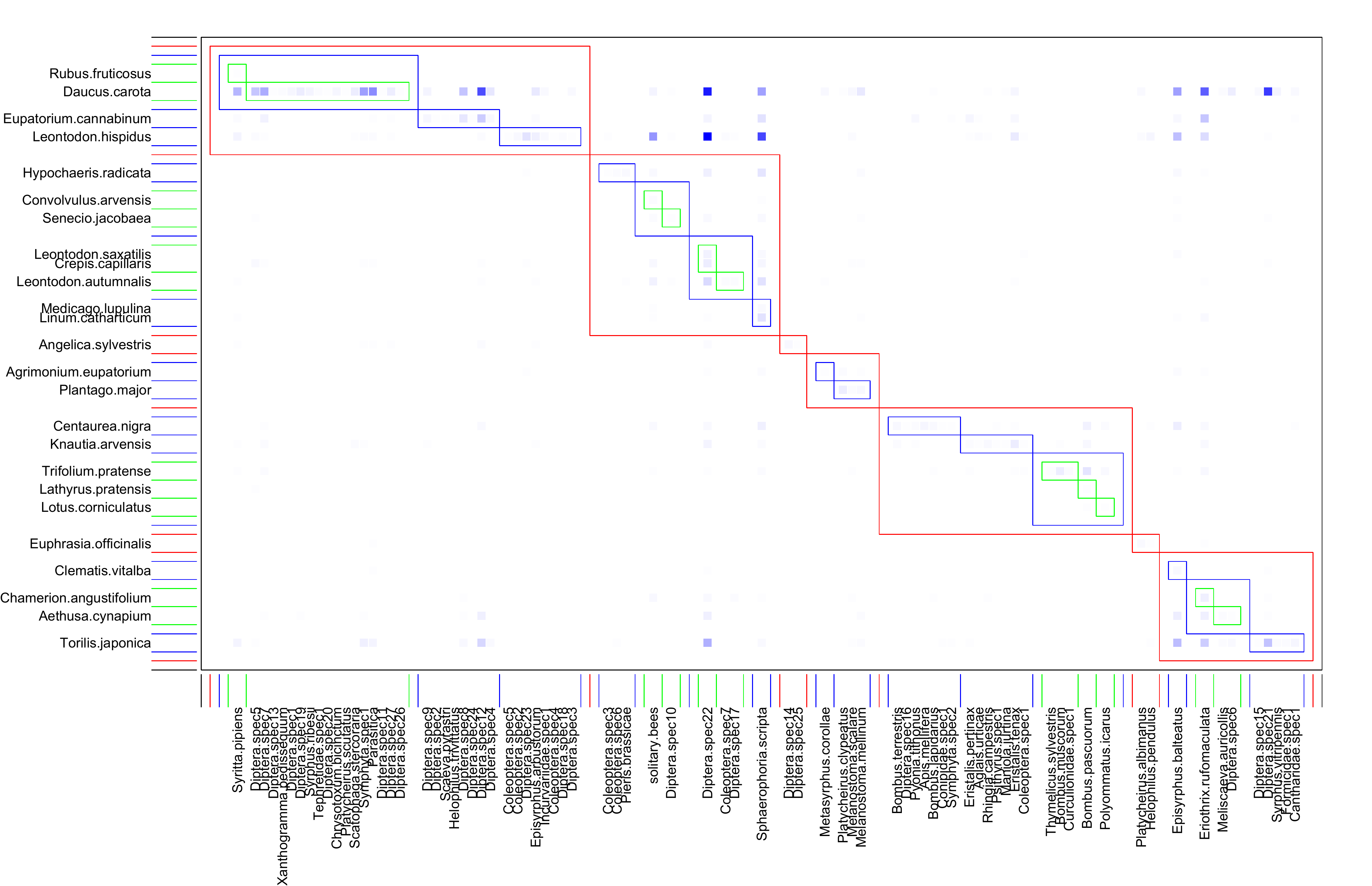}
	\caption{Interaction matrix featuring modules for the data of \citet{Memmott1999}. \emph{Top}: Modules identified by QuaBiMo (with steps=1E10, running for several hours; $Q=0.30$). Darker squares indicate more observed interactions. Red boxes delineate the seven modules. (Note that results may vary between runs.) In the central module yellow Asteraceae feature heavily, while a possible ecological cause pattern for the other modules is less apparent. \emph{Bottom}: Nested modules based on a recursive call of QuaBiMo. Module arrangement is slightly different from top, since the algorithm is stochastic.}
\label{fig:memmottmodules}
\end{figure}

As a typical analysis we shall use the relatively large (25 $\times$ 79) and well-sampled pollination network of \citet{Memmott1999}, which is provided along with the bipartite package:
\begin{verbatim}
> library(bipartite)
> mod <- computeModules(web=memmott1999, steps=1E8)
\end{verbatim}
The evaluation of these two lines will usually take about one minute and perform around 20 million MCMC moves. The resulting object stores the module composition and the likelihood of the solution found. The modularity value $Q$ of this solution is simply the likelihood value (0.18, this value may vary between runs; random seeding is not supported):
\begin{verbatim}
> mod@likelihood
[1] 0.18
\end{verbatim}
We can now plot the resulting modules to visualise the compartments (Fig.~\ref{fig:memmottmodules} top).
\begin{verbatim}
> plotModuleWeb(mod)
\end{verbatim}
To identify nested modules, we choose a lower value for steps (to reduce computation time), thus also yielding a different module structure at the highest level. Modularity value $Q$ will still be based on the non-recursive algorithm.
\begin{verbatim}
> modn <- computeModules(memmott1999, steps=1E6, deep=T)
\end{verbatim}
To be able to ecologically interpret these modules (Fig.~\ref{fig:memmottmodules} bottom), expert knowledge on the system is required. The computation of modularity is primarily an explorative tool helping the user to objectively detect pattern in typically noisy network data.

\section{Modularity \emph{Q} as a network index}\label{sec:session}
\begin{figure}
	\centering
	\includegraphics[width=0.5\textwidth]{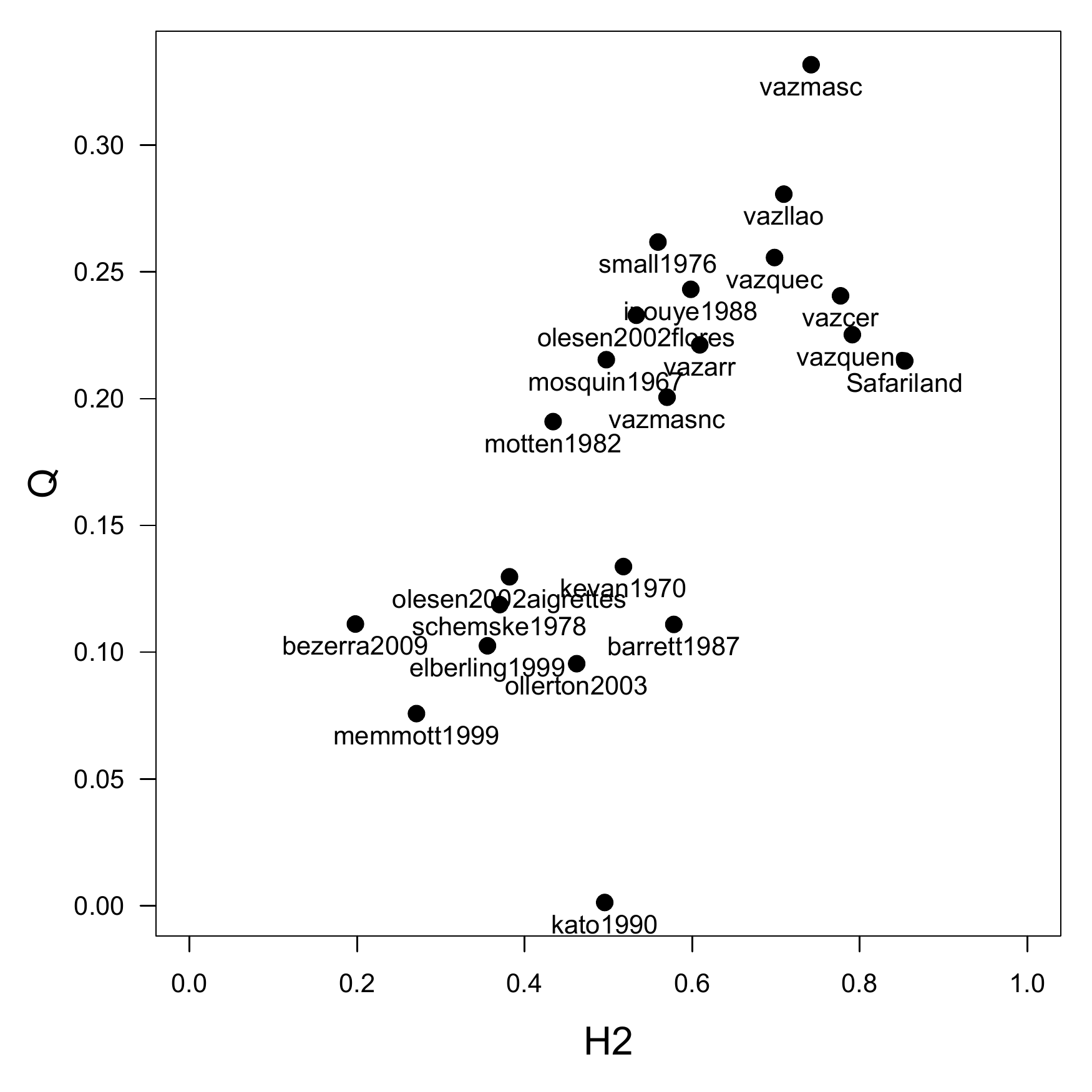}
	\caption{Modularity ($Q$) is highly correlated with specialisation $H_2'$ \citep{Bluthgen2006} across $22$ pollination networks. Names refer to network data sets in \texttt{bipartite} which were taken from \url{http://www.nceas.ucsb.edu/interactionweb}.}
\label{fig:QvsH2}
\end{figure}

Modularity $Q$ is likely to be correlated with other network metric, as specialisation of module members is the prime reason for the existence of modules. Across the $22$ quantitative pollination networks of the NCEAS ``interaction webs'' data base (\url{http://www.nceas.ucsb.edu/interactionweb}), $Q$ was evidently highly positively correlated with complementary specialisation $H_2'$ (Fig.~\ref{fig:QvsH2}).
Ecologically, the correlation with specialisation makes good sense. Modules only exist because some species do not interact with some others, i.e. because they are specialised. An overall low degree of specialisation is equivalent to random interactions, which will yield no modules. 

Furthermore, the absolute value of $Q$ \citep[like all network indices:][]{Dormann2009} is dependent on network size (i.e. the number of species) as well as the number of links and the total number of interactions observed \citep[see also][]{Thebault2013}. We would thus recommend a null model comparison \citep[e.g.][]{Vazquez2003a,Bluthgen2008,Dormann2009} to correct the observed value of $Q$ by null model expectation (e.g.~by standardising them to $z$-scores:
$z_{Q} = \frac{Q_{\text{observed}} - \overline{Q}_{\text{null}}}{\sigma_{Q_{\text{null}}}}$).
In R, this could be achieved by the following code (which will take more than one hour since we are computing modules in $100$ null model networks):
\begin{verbatim}
> nulls <- nullmodel(memmott1999, N=100, method="r2d")
> modules.nulls <- sapply(nulls, computeModules)
> like.nulls <- sapply(modules.nulls, function(x) x@likelihood)
> (z <- (mod@likelihood - mean(like.nulls))/sd(like.nulls))

[1] 7.088665
\end{verbatim}
This means that the observed modularity is $7$ standard deviations higher than would be expected from random networks with the same marginal totals (representing abundance distributions of plants and pollinators). Since $z$-scores are assumed to be normally distributed, values above $\approx 2$ are considered significantly modular. 


\subsection{Using modularity to identifying species with important roles in the network}
\citet{Guimera2005} and \citet{Olesen2007} propose to compute standardised connection and participation values, called $c$ and $z$, for each species to describe their role in networks, where $c$ refers to the between-modules connectivity \citep[called ``participation coefficient'' $P$ by][]{Guimera2005} and $z$ refers to within-module degrees. Both are computed on the number of links and are not weighted by the number of interactions per link. \citet{Guimera2005} suggest critical values of $c$ and $z$ of $0.625$ and $2.5$, respectively. Species exceeding both of these values are called ``hubs'' because they link different modules, combining high between- with high within-module connectivity. 

\begin{figure}
\hfill
\includegraphics[width=0.48\textwidth]{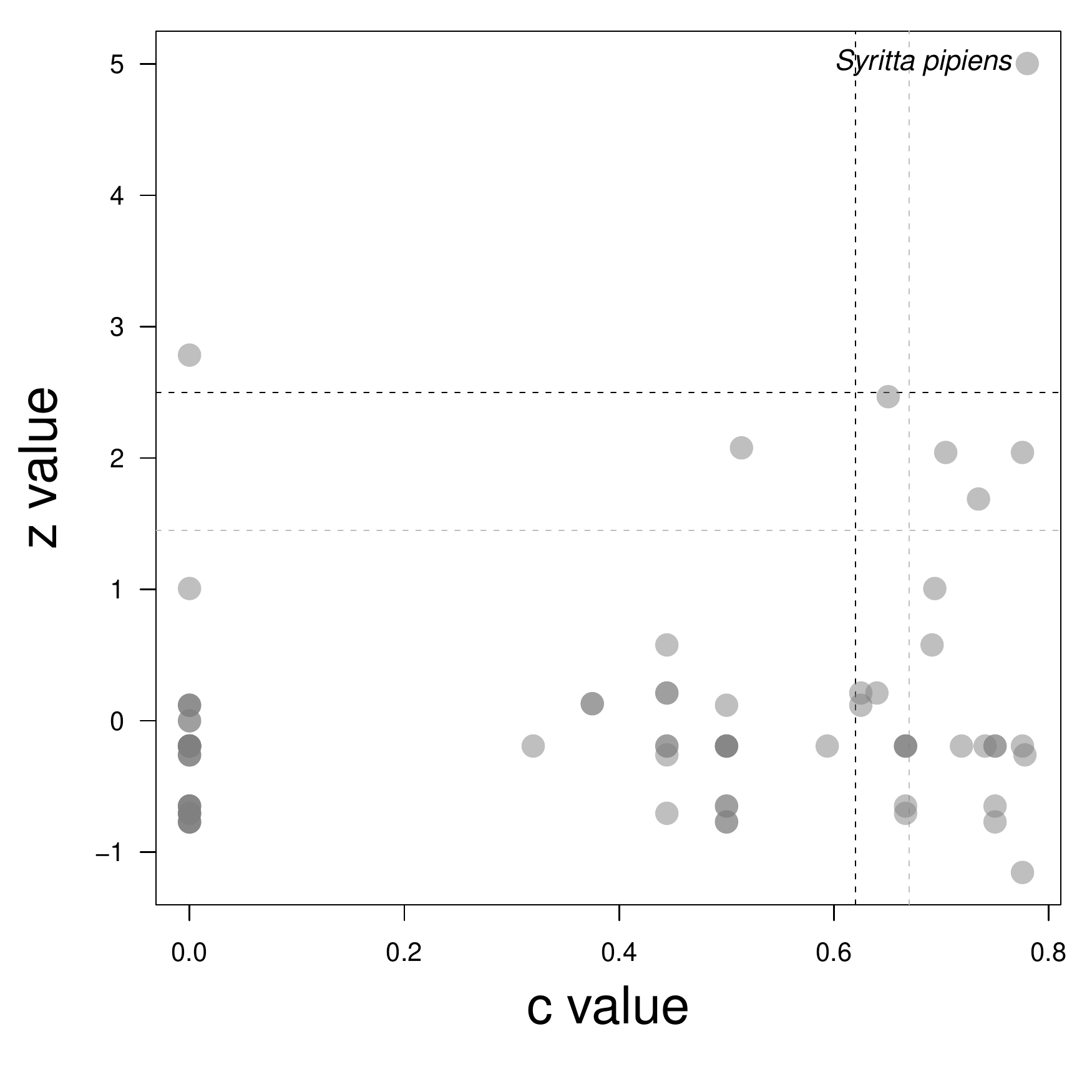}
\hfill
\includegraphics[width=0.48\textwidth]{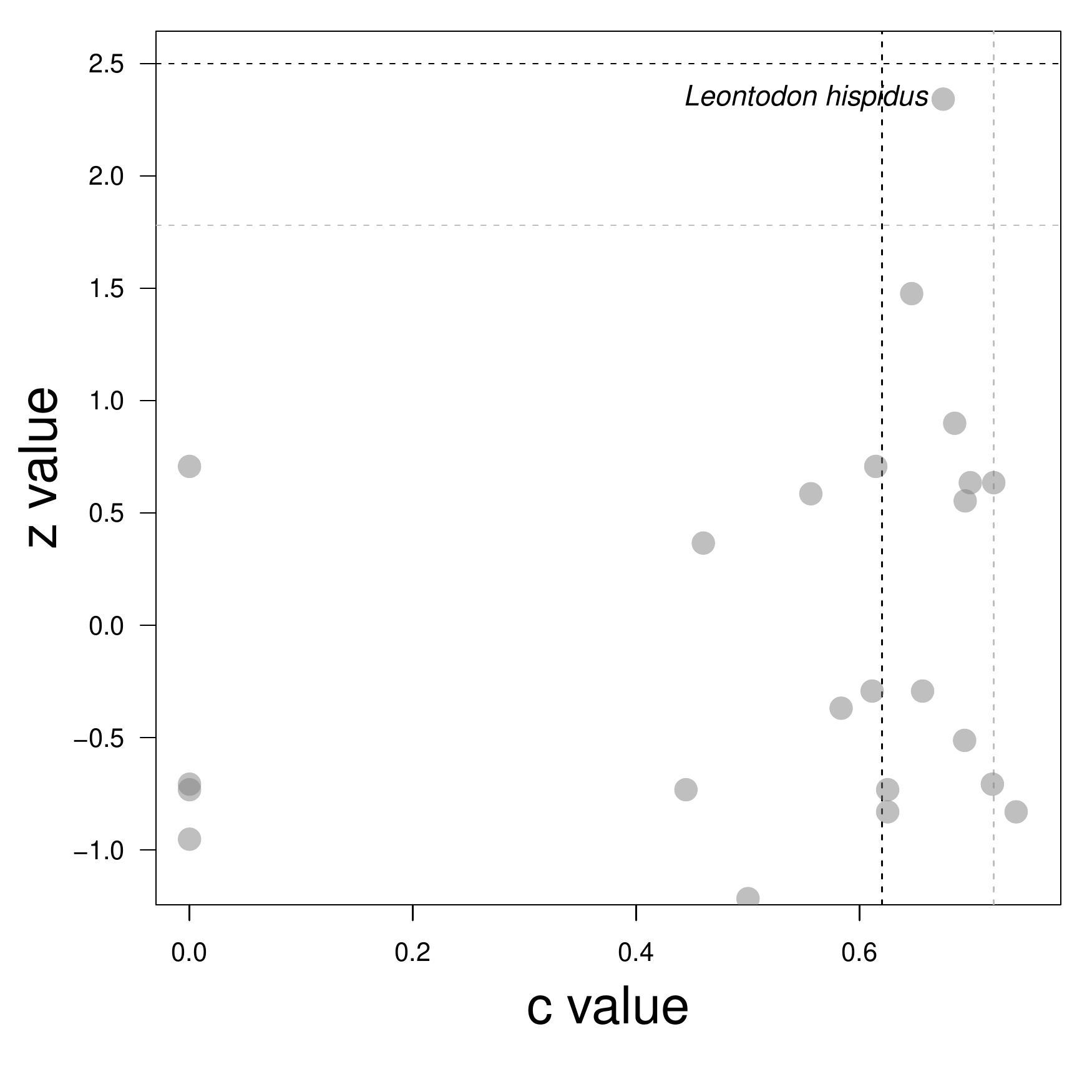}
\hfill
\caption{Connection ($c$) and participation ($z$) values for pollinators (left) and plants (right) in the network of \citet{Memmott1999}. Dashed black lines indicate critical values according to \citet{Olesen2007}, those in grey $95$\% quantiles from $100$ null models (see text).}
\label{fig:czvalues}
\end{figure}

In the case of the pollination network of Fig.~\ref{fig:bipartitegraph}, $c$-values range between $0$ and $0.78$ (with $23$ of $79$ pollinators and $13$ of $25$ plant species exceeding the threshold of $0.625$); $z$-values range between $-1.21$ and $5.00$ (with two pollinators but no plant species exceeding the value of $2.5$: Fig.~\ref{fig:czvalues}). Put together, only the syrphid \emph{Syritta pipiens} (and hawkbit \emph{Leontodon hispidus} almost) exceeded both thresholds and would thus be called a ``hub species''. As can be seen in Fig.~\ref{fig:memmottmodules}, this syrphid is relatively rare but clearly not randomly distributed over the six modules, thus linking modules three, five and six (from the left). In contrast, \emph{Leontodon hispidus} is a common plant species, visited by many different pollinators, and it actually links all modules with the exception of module two. 

To objectively define this threshold one could run null models of the original network and employ $95$\% quantiles as critical $c$- and $z$-values. For the pollinators in the network of Fig.~\ref{fig:bipartitegraph} these would be $0.67$ ($\pm 0.039$) and $1.45$ ($\pm 0.220$), respectively, based on $100$ null models (for the plants: $c_\text{critical} = 0.72 \pm 0.036$ and $z_\text{critical} = 1.78 \pm 0.297$; Fig.~\ref{fig:czvalues} left). While this has little effect for plant species (except for moving \emph{Leontodon hispidus} across the threshold), three more pollinators would become hub species (the common hoverfly \emph{Episyrphus balteatus}, the tachinid fly \emph{Eriothrix rufomaculata} and undetermined fly ``\emph{Diptera} spec.22'').

\section{Conclusion}
We here presented an algorithm to compute modularity $Q$ and detect modules in weighted, bipartite networks. In a preliminary analysis, this approach was able to identify meaningful ecological modules in frugivore networks (Schleuning \emph{et al}., in submission). Because it uses the strength of links as quantitative information, this approach should be much more sensitive, and also more specific, than current binary algorithms. By making the algorithm easily available we hope that network ecology will benefit from new insights into the structure of interaction networks.

\section{Acknowledgements}
We like to thank Aaron Clauset for inventing the hierarchical random graph idea and for making his code freely available. CFD acknowledges funding by the Helmholtz Association (VH-NG 247). Many thanks to Lili Ingmann and Matthias Schleuning for extensively testing the robustness and ecological meaningfulness of this algorithm.

\bibliographystyle{mee}
\bibliography{../../../../../../LiteratureOrganisation/CFD_library}

\section*{Appendix A: Formal definition of the identification of module vertices}

Consider an edge $(i,j) \in E$ with weight $w_{ij}$ representing the strength of interaction between vertices $i$ and $j$. In a bipartite graph $\overline{G}$ maintaining for each vertex its original sum of edge weights, but disregarding the modular structure of $G$, the weight $\overline{w}_{ij}$ of the edge between vertices $i$ and $j$ is given by
\begin{equation}
\overline{w}_{ij} = \left\{\begin{array}{cl} \displaystyle\sum{w_{i.}} \times \sum{w_{.j}}\ , & \mbox{if }{ i \in V_A \Leftrightarrow j \in V_B}\\\\ 0, & \mbox{else.} \end{array}\right.
\end{equation}
Therefore, the difference of edge weight and expected edge weight
\begin{equation}
w'_{ij} = w_{ij} - \overline{w}_{ij}
\end{equation}
is positive, if within module, and negative, if outside module.\\


Therefore, the algorithm attempts to find the best trade-off between a maximum sum of $w'$ within  modules and a minimum sum of $w'$ outside.

Given a division of $V$ into a set of non-overlapping subgraphs $C$, we define 
\begin{equation}
{g(C)} = \left\{\begin{array}{cl} \displaystyle\sum_{i \in V_A}{\sum_{j \in V_B}{\delta_C(i,j) \times w'_{ij} - (1 - \delta_C(i,j)) \times w'_{ij}}}\ , & \mbox{if }{\forall c\in C}{\textrm {: c\ is\ connected\ graph}}\\\\\displaystyle\ -\infty, & \mbox{else,}\end{array}\right.
\end{equation}
where
\begin{equation}
\delta_C(i,j) = \left\{\begin{array}{cl} 1, & \mbox{if }{i \in c \wedge j \in c \wedge c \in C}\\\\ 0, & \mbox{else.} \end{array}\right.
\end{equation}
Obviously, $g(C)$ has to be maximized in order to find the best division of $V$ into modules $C$. For achieving this goal, we modify the algorithm of \citet{Clauset2008}.

\bigskip 

Let $D$ be a binary tree with arbitrarily connected internal vertices $v \in V_\text{intern}$ and with $n$ leaves representing the vertices of $G$ and initially arranged in an arbitrary order.
A module $c$ within $D$ is defined as the set of leaves of the sub-tree rooted at an internal vertex $v$ meeting following requirements:
\renewcommand{\labelenumi}{\Roman{enumi}}
\begin{enumerate}
\item $v$ has at least one child being a leaf.
\item No ancestor of $v$ has a child being a leaf.
\item $c \cap V_A \neq \emptyset \wedge c \cap V_B \neq \emptyset$, i.e. there is at least one vertex $v_A \in V_A$ and at least one vertex $v_B \in V_B$ within $c$.
\end{enumerate}
Due to requirement I it is obvious that there are at most $min(|V_A|,|V_B|)$ modules. Note that due to requirement II on each path from the root of $D$ to a leaf there is exactly one internal vertex shaping a module. For convenience, we will use the term 'module vertex' for this kind of vertex.

Each internal vertex $v$ is assigned the information $r_v$ whether it is the root of a sub-tree of $D$ representing a module or whether it is below or above such an internal vertex. Let $r_v = 1$ if $v$ is above a module vertex, $r_v = 0$ if $v$ is a module vertex itself and let $r_v = -1$ if $v$ is below a module vertex.

Additionally, each internal vertex $v$ is assigned its contribution $g_v$ to $g(C)$
\begin{equation}
g_v = \left\{\begin{array}{cl} \displaystyle\ +\ \sum_{i \in {\cal L}_v}{\sum_{j \in {\cal R}_v}{w'_{ij}}}\ , & \mbox{if }{r_v \leq 0\ \wedge\ \displaystyle\sum_{i \in {\cal L}_v}{\sum_{j \in {\cal R}_v}{w_{ij} > 0}}}\\\\  \displaystyle\ -\ \sum_{i \in {\cal L}_v}{\sum_{j \in {\cal R}_v}{w'_{ij}}}\ , & \mbox{if }{r_v = 1} \\\\\displaystyle\ -\infty, & \mbox{else,}\end{array}\right.
\end{equation}
where ${\cal L}_v$ is the set of leaves of the sub-tree rooted at the left child of $v$ and, analogously, ${\cal R}_v$ is the set of leaves of the sub-tree rooted at the right child of $v$.

For $C$ given by the current state of $D$, $g(C)$ can now be rewritten as
\begin{equation}
{g(C)} = \displaystyle\sum_{v \in V_{intern}}{g_v}\hspace{10 mm}.
\end{equation}
In order to compute max$(g(C))$, the subtrees of $D$ have to be re-arranged. The algorithm therefore randomly selects an edge $e$ of $D$ connecting two internal vertices $v_i$ and $v_j$. Let w.l.o.g. $e$ be the left edge of $v_j$ connecting it to its child $v_i$. Then there are three subtrees ${\cal L}_{v_i}$, ${\cal R}_{v_i}$ and ${\cal R}_{v_j}$ originating from $v_i$ and $v_j$, respectively, and two possible rearrangements $\alpha$ and $\beta$ (Fig. \ref{fig:swaps}) of which one is chosen randomly and simulated. In re-arrangement $\alpha$, sub-trees ${\cal R}_{v_i}$ and ${\cal R}_{v_j}$ are permuted, in rearrangement $\beta$ sub-trees ${\cal L}_{v_i}$ and ${\cal R}_{v_j}$. The change $dg$ in $g(C)$ resulting from the rearrangement is computed according to $r_{v_i}$ and $r_{v_j}$.

\end{document}